\documentstyle[prb,preprint,aps]{revtex}

\begin{document}
\draft
\title{Stochastic resonance induced by random fields 
in ferroelectrics }
\author{Zhi-Rong Liu
\footnote{Electronic address: zrliu@phys.tsinghua.edu.cn} 
and Wenhui Duan}
\address{Department of Physics, Tsinghua University, 
Beijing 100084, People's Republic of China}
\author{Bing-Lin Gu} 
\address{Center for Advanced Study, and Department of Physics, 
Tsinghua University, Beijing 100084, People's Republic of China}
\author{Xiao-Wen Zhang}
\address{State Key Laboratory of New Ceramics and Fine Processing, 
Department of Materials Science and Engineering, 
Tsinghua University, Beijing 100084, People's Republic of China}
\maketitle

\begin{abstract}
We propose a mechanism of stochastic resonance induced by spatial noise 
(electric random fields) in ferroelectrics. 
The calculations demonstrate the 
characteristic of the stochastic resonance:  
certain random field can increase the susceptibility, the 
phonon occupancy, and the overcooled temperature. This mechanics of 
stochastic resonance may be general in other disordered systems.

\end{abstract}

\pacs{PACS:02.50.-r 05.40.-a 77.80.-e 77.22.Ch}

\vspace{2mm}


Stochastic resonance (SR) is a nonlinear phenomenon in which 
the degree of order rather than disorder can be enhanced by 
the presence of optimized random noise.\cite{1} 
It was originally proposed to explain the periodicity in the recurrences 
of the Earth's ice ages as the result of stochastic and weak periodic 
forces acting in concert on a bistable global climate model.\cite{3} 
Experimentally, SR was first demonstrated in a Schmitt trigger circuit 
and a bistable ring-laser system.\cite{5} Since then stochastic 
resonance has been reported in various systems from crayfish to SQUID 
and spin system.\cite{7,8} Variations and extensions of the classical 
definition of SR were also developed to include aperiodic (e.g., dc or 
wideband) signals. 
SR is particularly interesting
for neurobiological systems because it provides a
mechanism for such systems to detect and process weak
signals.\cite{7,10} In applications, SR may be useful in physical, 
technological and biomedical contexts.

Stochastic resonance usually occurs in systems with multiple stable 
states. On the other hand, ferroelectrics are known as multistable 
systems since 
the polarization can emerge along more than one directions. In the 
so called relaxor ferroelectrics (relaxors) such as 
Pb(Mg$_{1/3}$Nb$_{2/3}$)O$_3$ (PMN), there exists spatial random electric 
fields induced by point charge defects, nano-domain textures, and 
the direct interactions of polar ions, etc.\cite{13} 
Some specific characteristics of the relaxors were interpreted in terms of different
models adopting random fields.\cite{16} Recently, it 
was shown that an optimal intensity of random fields can promote 
the spontaneous appearance of a first-order phase transition by 
increasing the overcooled temperature of ferroelectrics.\cite{19} 
Well then a question naturally arises: {\it Can the random fields in 
ferroelectrics produce a certain stochastic resonance?}

To resolve the above problem, in this paper 
we investigate a ferroelectric system 
with random fields, and show that a stochastic resonance does 
occur in this case.

The basic picture of stochastic resonance is generally illustrated 
using the following simple example:\cite{1,20} Imagine a particle subject to 
friction, moving in a double-well potential. A weak periodic signal 
force is applied on the particle (here, 
``weak'' means that the signal force is far too small to excite the 
particle over the barrier between the wells). 
Meanwhile, random noise is applied to tilt the potential and could result
in irregular switching of the particle between the wells. 
With the help of certain noise,
the weak periodic signal can effectively 
make the particle hop over the barrier, and 
produce the maximal signal-to-noise ratio (SNR) at a nonzero level of input noise. 
This phenomenon is called stochastic resonance. 
In general, it is assumed that the signal period is longer 
than some characteristic intrawell relaxation time for the system. 
According to the statistical mechanics, the time average of a mechanical 
quantity of a system is equal to the ensemble average (ergodic 
hypothesis). So the above stochastic resonance can be constructed in 
another way: many particles moving in different double-well potentials, 
respectively. In this case, the applied noise does not change in time, but changes 
in space, i.e., double-well potentials tilt in random slopes. This 
is just the physical picture of the ferroelectric system with random 
fields where the tilt of potential is caused by random  electric fields.\cite{21} 
Therefore, a stochastic resonance with ``spatially extended" noise can be 
reasonably expected in ferroelectrics.

For the ferroelectrics with the second-order phase transition, it 
was shown that the critical temperature monotonously declines with 
increasing random fields.\cite{21} In this case, even if a phenomenon
of stochastic resonance appears, 
it seems not to lead to interesting practical applications.
Therefore, in this paper, we focus our study on the 
ferroelectrics with the first-order phase transition, i.e., 
with a triple-well potential (See Fig. 1). For this system, when the 
particle (polar ion) jumps from the middle well to the sideward wells, 
a paraelectrics-to-ferroelectrics phase transition takes place. Since 
optimal noise can enhance the hopping of particle between wells, a 
random-field-induced ferroelectric transition and nonmonotonic behaviors 
might be observed. 

A microscopic 
Hamiltonian describing order-disorder  ferroelectrics is introduced as\cite{19}:
\begin{equation}
H=\left[
\begin{array}{cccccccc}
-Ep_0 & \frac{\Omega}{2} & 0 & 0 & 0  & 0 & 0 & 0\\
\frac{\Omega}{2} & Ep_0 & 0 & 0 & 0  & 0 & 0 & 0\\
0 & 0 & -\frac{1}{3}Ep_0 & \frac{\Omega}{2} & 0 & 0 & 0  & 0 \\
0 & 0 & \frac{\Omega}{2} & \frac{1}{3}Ep_0 & 0 & 0 & 0  & 0 \\
0 & 0 & 0 & 0 & -\frac{1}{3}Ep_0 & \frac{\Omega}{2} & 0 & 0 \\
0 & 0 & 0 & 0 & \frac{\Omega}{2} & \frac{1}{3}Ep_0 & 0 & 0 \\
0 & 0 & 0 & 0 & 0 & 0 & -\frac{1}{3}Ep_0 & \frac{\Omega}{2} \\
0 & 0 & 0 & 0 & 0 & 0 & \frac{\Omega}{2} & \frac{1}{3}Ep_0  
\end{array} \right],
\end{equation}
where $E$ is the equivalent field, and $p_0$ is the dipole 
moment of the ion in a local polar position. $\Omega$ 
is the tunneling frequency between wells. Under a mean field 
approximation, 
\begin{equation}
Ep_0=J\frac{\langle p \rangle}{p_0}+E_{\rm ext}p_0+E_{\rm rand}p_0,
\end{equation}
where $\langle p \rangle$ is the thermal average of the dipole 
moment, and $J$ is the coupling energy between polar ions. $E_{\rm ext}$ is the 
applied external electric field, and $E_{\rm rand}$ is the internal random field 
in the system which is assumed to have a Gaussian distribution with a width $\sigma_e$. 
Hamiltonian in Eq. (1) produces a second-order phase transition for 
$\Omega<0.56J$ and a first-order one for $0.56J<\Omega<2J$ when there 
is no random field.\cite{19}   

From Eq. (2) one can see that the uniform variable $E_{\rm ext}$ is 
the input signal, and the inhomogeneous $E_{\rm rand}$ is the input random noise. 
The average polarization $\langle p \rangle$ is the output signal, and  
appears in the right hand of Eq. (2) as the feedback input through the 
ferroelectric interaction between polar ions. When there is no external 
input ($E_{\rm ext}=0$), the feedback mechanism can preserve a nonzero 
output at low temperatures and the system stays in a ferroelectric state. 
The over-cooled temperature $T_{-}$ is used to measure the spontaneous 
phase transition in a cooling process. In Fig. 2, $T_{-}$ 
is given as function of the intensity 
of noise ($\sigma_e$) when $\Omega=0.63J$. 
It can be seen that $T_{-}$ increases first and then declines with 
increasing random field, which is consistant with previous calculation.\cite{19}
 $T_{-}$ reaches its maximum at  
$\sigma_e=0.23J/p_0$. This means that a nonzero random field can induce the 
spontaneous appearance of the ferroelectric state, i.e., enhance 
the output signal. Stochastic resonance really exists in the ferroelectrics.

When the temperature is higher than the maximum over-cooled temperature, 
the system stays in a paraelectric state for any noise. The 
effect of random fields can then be understood by investigating the 
response of the system to a weak external field. The ratio of the output 
and input signals is measured by the dielectric susceptibility, 
\begin{equation}
\chi=\frac{P}{\varepsilon_0 E_{ext}}
     =\frac{N_0 \langle p \rangle}{\varepsilon_0 E_{ext}},
\end{equation}
where $P$ is the polarization, and $N_0$ is the number of polar ions 
in unit volume. The calculation result is depicted in Fig. 3. 
It shows that appropriate noise can 
{\it greatly enhances} the output signal and results 
in a peak of the susceptibility, 
which is just the characteristic signature of stochastic resonance.

In the paraelectric state with no external field, the polarization 
does not keep zero but fluctuates due to the thermal movement. 
The polarization fluctuation can be described in terms of the 
ferroelectric soft mode and the phonon.
The dependence of the soft-mode 
frequency and the phonon occupancy on the random 
fields is demonstrated in Fig. 4. Nonmonotonic behaviors are observed for  
both quantities. Stochastic resonance 
causes the decrease of the soft-mode frequency and the increase of the phonon 
occupancy, i.e., makes it easier for the thermal fluctuation of polarization 
to take place.

For ferroelectrics, the random fields are mainly determined by the impurity and the microstructure 
imhomogeneity in the system. If there are more charge impurities or the 
microstructure is more inhomogeneous, the random fields are stronger. 
The features of imhomogeneity include grain size, porosity, grain boundary, 
chemical imhomogeneity, and domain/twin structure, etc. The imhomogeneity can 
be controlled by thermal treatments in some degrees. So the stochastic 
resonance can be realized through the doping and thermal treating. Further 
experiments in this aspect are needed to testify the theory.

In summary, we have demonstrated a specific stochastic resonance in 
ferroelectric systems, which is not produced by the temporal noise, but by 
the spatial noise (random electric fields). The characteristics of stochastic 
resonance are revealed via the calculations of the over-cooled temperature, 
the dielectric susceptibility, the soft-mode frequency, and the phonon 
occupancy, which all show nonmonotonic dependence on the noise. 
The spatial stochastic resonance may be 
useful in materials applications as well as in theoretical interests. For example, 
the stability of paraelectric and 
ferroelectric phases can be adjusted by modulating the over-cooled temperature, 
and the dielectric susceptibility of materials can be increased by selecting 
appropriate amount of random fields. The 
mechanics of the spatial stochastic resonance revealed in this paper 
might also be applicable in a number of 
inhomogeneous systems whose properties are not well understood.


This work was supported by the Chinese National Science Foundation 
(Grant NO. 59995520) and State Key Program of Basic Research 
Development (Grant No. G2000067108).

\begin{figure}[tbp]
\caption{Schematic graphics of a treble-well potential and the 
hopping of a particle.}
\end{figure}

\begin{figure}[tbp]
\caption{Over-cooled temperature (in unit of $J/k_B$) of the ferroelectrics 
described by the Hamiltonian in Eq. (1) as a function 
of the random field width (in units of $J/p_0$) when $\Omega=0.63J$.  }
\end{figure}

\begin{figure}[tbp]
\caption{Dielectric susceptibility $\chi$ as a function of the random 
field width (in units of $J/p_0$) when $\Omega=0.63J$ and 
$k_BT=0.2J$. $\chi$ is normalized by a factor 
$N_0 p_0^2/\varepsilon_0 J$. }
\end{figure}

\begin{figure}[tbp]
\caption{Random field width dependence of the ferroelectric phonon occupancy 
when $\Omega=0.63J$ and $k_BT=0.2J$. Inserted graphics is the soft-mode 
frequency. }
\end{figure}


\begin{references}
\bibitem{1} K. Wiesenfeld and F. Moss, Nature (London) {\bf 373}, 33 (1995);
 L. Gammaitoni, P. Hanggi, P. Jung, and F. Marchesoni, 
Rev. Mod. Phys. {\bf 70}, 223 (1998). 

\bibitem{3} R. Benzi, S. Sutera, and A. Vulpiani, J. Phys. A {\bf 14}, 
L453 (1981); C. Nicolis, Tellus {\bf 34}, 1 (1982).

\bibitem{5} S. Fauve, and F. Heslot, Phys. Lett. A {\bf 97}, 5 (1983);
 B. McNamara, K. Wiesenfeld, and R. Roy, Phys. Rev. Lett. 
{\bf 60}, 2626 (1988).

\bibitem{7} J. K. Douglass, L. Wilkens, E. Pantazelou, and F. Moss, 
Nature (London) {\bf 365}, 337 (1993).

\bibitem{8} A. D. Hibbs, A. L. Singsaas, E. W. Jacobs, A. R. Bulsara, 
and J. J. Bekkedahl, J. Appl. Phys. {\bf 77}, 2582 (1995);
 D. Gourier and D. Gerbault, Phys. Rev. B. {\bf 57}, 
2679 (1998).

\bibitem{10} J. J. Collins, C. C. Chow, and T. T. Imhoff, 
Nature (London) {\bf 376}, 236 (1995);
 J. E. Levin and J. P. Miller, Nature (London) {\bf 380}, 
165 (1996); D. Nozaki, D. J. Mar, P. Grigg, and J. J. Collins, 
Phys. Rev. Lett. {\bf 82}, 2402 (1999).

\bibitem{13} L. E. Cross, Ferroelectrics {\bf 76}, 241 (1987);
 Z. G. Ye, Key Engineering Materials {\bf 155-156}, 81 (1998);
 H. Qian and L. A. Bursill, Int. J. Mod. Phys. B 
{\bf 10}, 2027 (1996).

\bibitem{16} D. Viehland, S. J. Jang, and L. E. Cross, J. Appl. 
Phys. {\bf 68}, 2916 (1990);
 H. Gui, B. L. Gu, and X. W. Zhang, Phys. Rev. B 
{\bf 52}, 3135 (1995);
 V. Westphal, W. Kleemann, and M. D. Glinchuk, 
Phys. Rev. Lett. {\bf 68}, 847 (1992).

\bibitem{19} Z. R. Liu, B. L. Gu, and X. W. Zhang, Appl. Phys. Lett. (in press); 
see also arXiv:cond-mat/0008398.

\bibitem{20} B. McNamara and K. Wiesenfeld, Phys. Rev. A 
{\bf 39}, 4854 (1989).

\bibitem{21} M. D. Glinchuk and V. A. Stephanovich, 
J. Phys.: Condens. Matter {\bf 6}, 6317 (1994).

\bibitem{22} V. Berdichevsky and M. Gitterman, Phys. Rev. E {\bf 60}, 
1494 (1999).

\end{references}
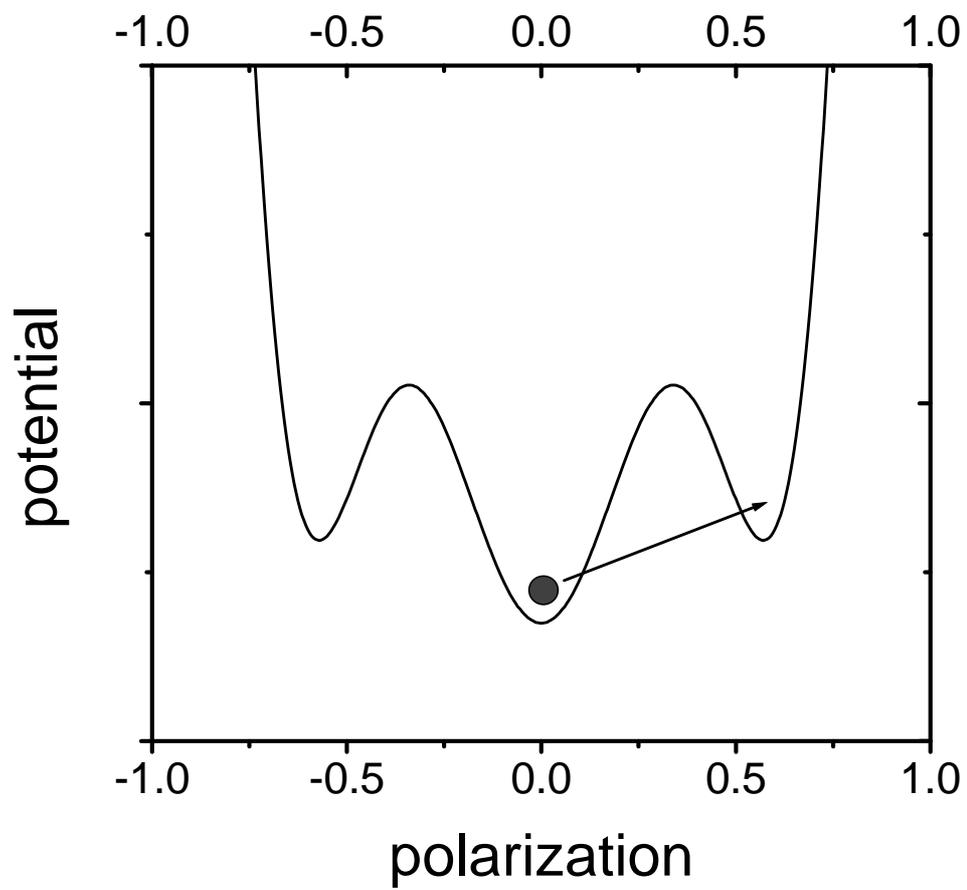
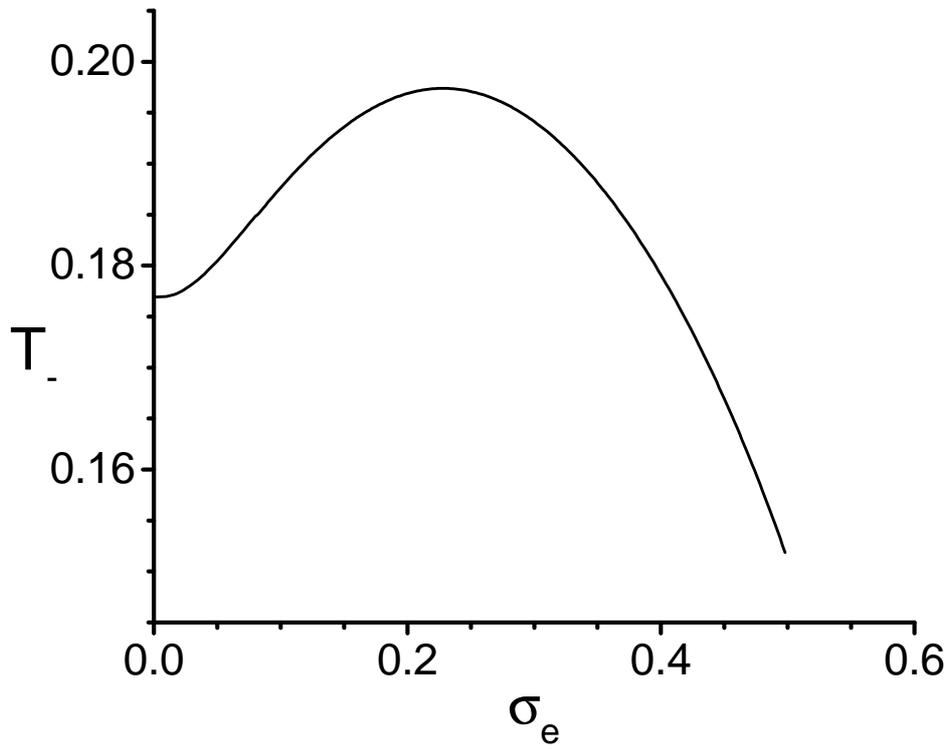
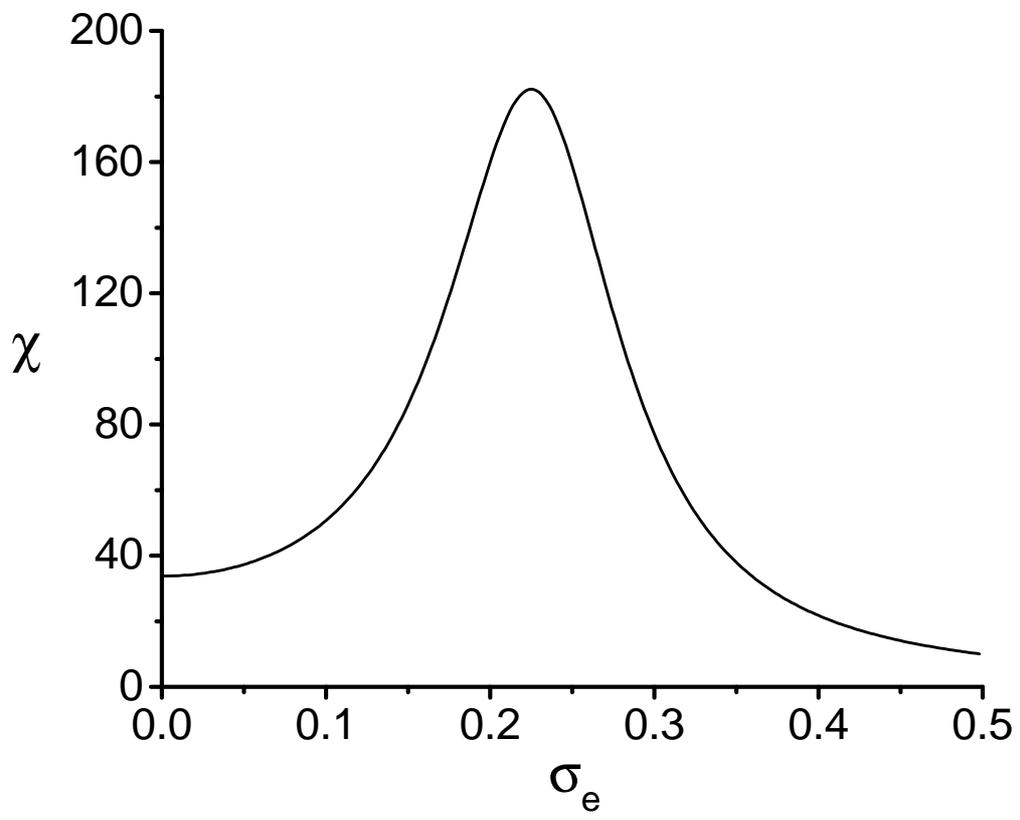
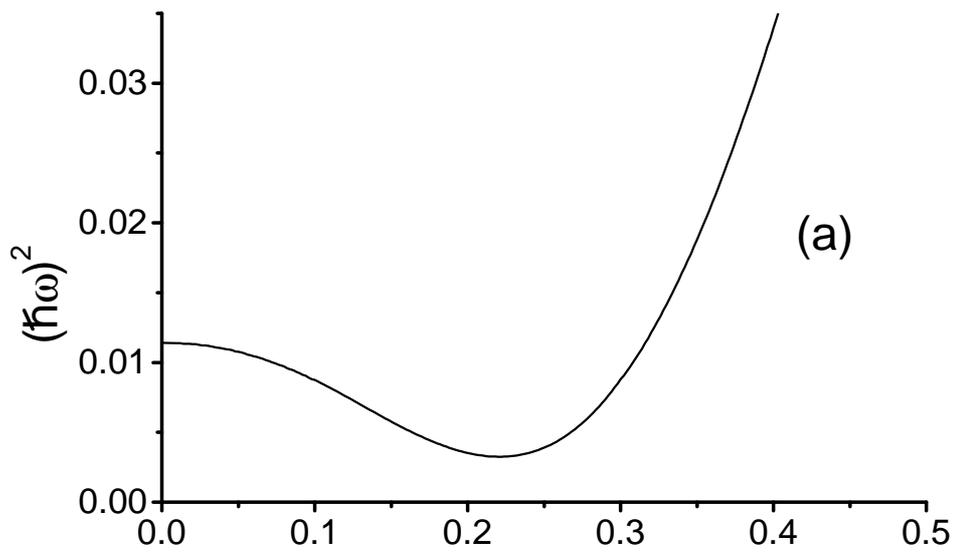
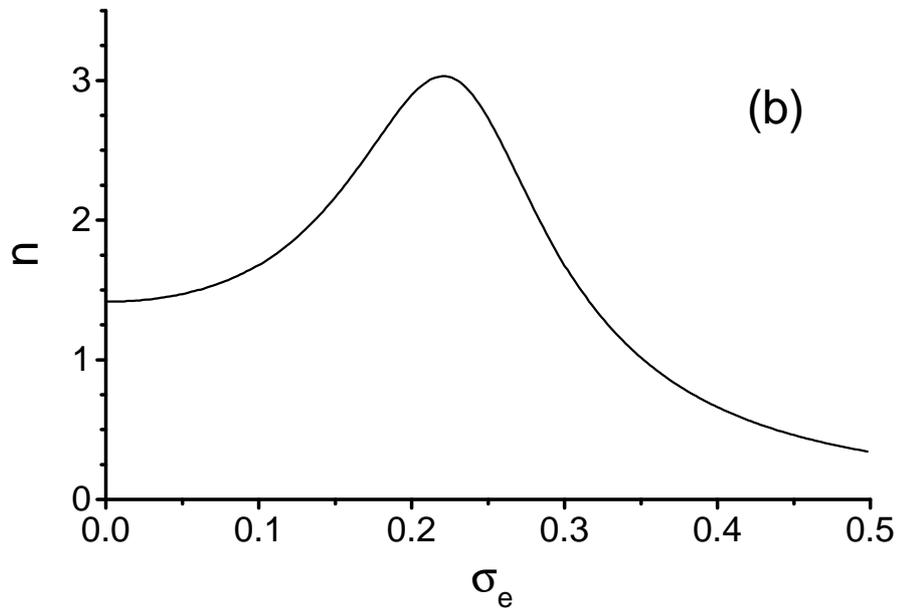
\end{document}